\documentclass{PoS}
\usepackage{graphicx}
\usepackage{caption}
\usepackage{subcaption}
\usepackage{amsmath}

\usepackage{cleveref}

\title{Supernova Neutrino Detection in the Deep Underground Neutrino Experiment}

\ShortTitle{Supernova Neutrino Detection in DUNE}

\author{\speaker{A. Gallego-Ros} \textbf{on behalf of the DUNE Collaboration}\\
        CIEMAT, Centro de Investigaciones Energ\'eticas, Medioambientales y Tecnol\'ogicas \\Madrid (Spain)\\
        E-mail: \email{ana.gallego@ciemat.es}}

\abstract{The Deep Underground Neutrino Experiment (DUNE) is a dual-site experiment for long-baseline neutrino oscillation studies and for neutrino astrophysics and nucleon decay searches. The Far Detector of DUNE will consist of four 10 kt liquid argon time-projection-chambers (LAr TPC) placed in the Sanford Underground Research Facility (SURF) at 1300 km distance from the neutrino beam at Fermilab. The underground location of the Far Detector, at 4300 m.w.e. depth, is essential to be able to study rare and low-energy processes. DUNE will have a unique sensitivity to the electron neutrino flavor component of the core collapse of a massive star. With a large mass DUNE will be able to detect core collapse events in the Milky Way and its neighborhood. The present document reviews some recent progress on detection and reconstruction of supernova burst neutrinos in DUNE, including the contribution of the light detection system.}

\FullConference{
European Physical Society Conference on High Energy Physics - EPS-HEP2019 -\\
			10-17 July, 2019\\
			Ghent, Belgium}

\begin{document}

\section{Introduction: what is DUNE?}

DUNE is a dual-site experiment set for operation in 2026 \cite{bib1}. The physics program of DUNE includes long-baseline neutrino oscillation studies, neutrino astrophysics and nucleon decay searches. The Far Detector (FD) will consist of four 10 kt fiducial mass liquid argon time-projection-chambers (LAr TPC) located in the Sanford Underground Research Facility (SURF) at 1300 km distance from the high purity muon neutrino beam produced at Fermilab. Each of the four modules will be either Single Phase (SP) or Dual Phase (DP). Several prototypes are underway at CERN to demonstrate these technologies at the large scale. When a charged particle crosses the detector, it produces free electrons and scintillation light. The combination of both the ionization charge and light readouts of the LAr TPC allows the 3D reconstruction of events with excellent spatial resolution.

\section{Neutrinos from a core-collapse supernova (SN)}

DUNE is expected to detect $\sim$3000 neutrinos from a SN at 10 kpc \cite{bib2}. This will contribute to a better understanding of the collapse dynamics and the SN properties. Three main phases can be distinguished in the SN neutrino burst:

\begin{itemize}
\item \textbf{Neutronization burst:} large electron neutrino emission with shock breakout information
\vspace{-0.3cm}
\item \textbf{Accretion phase:} different luminosity of electron and non-electron flavors
\vspace{-0.3cm}
\item \textbf{Cooling phase:} main part of the signal when proto-neutron star releases its trapped energy
\end{itemize}

\begin{figure}[b!]
\centering
\includegraphics[height=0.35\textwidth]{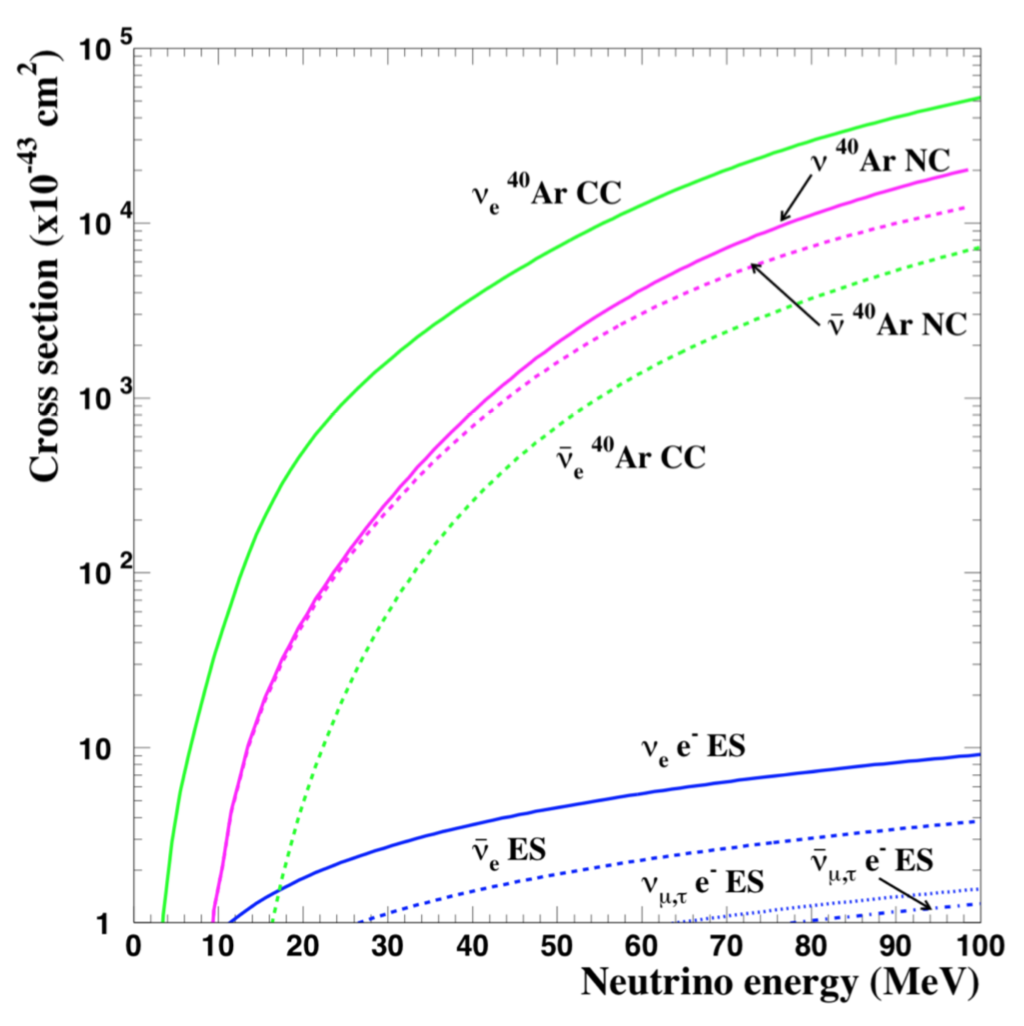}
\caption{Cross sections for the most relevant SN neutrino interactions in argon \cite{bib4}.}
\label{fig:cross}
\end{figure}

\begin{figure}[h!]
\centering
\includegraphics[height=0.3\textwidth]{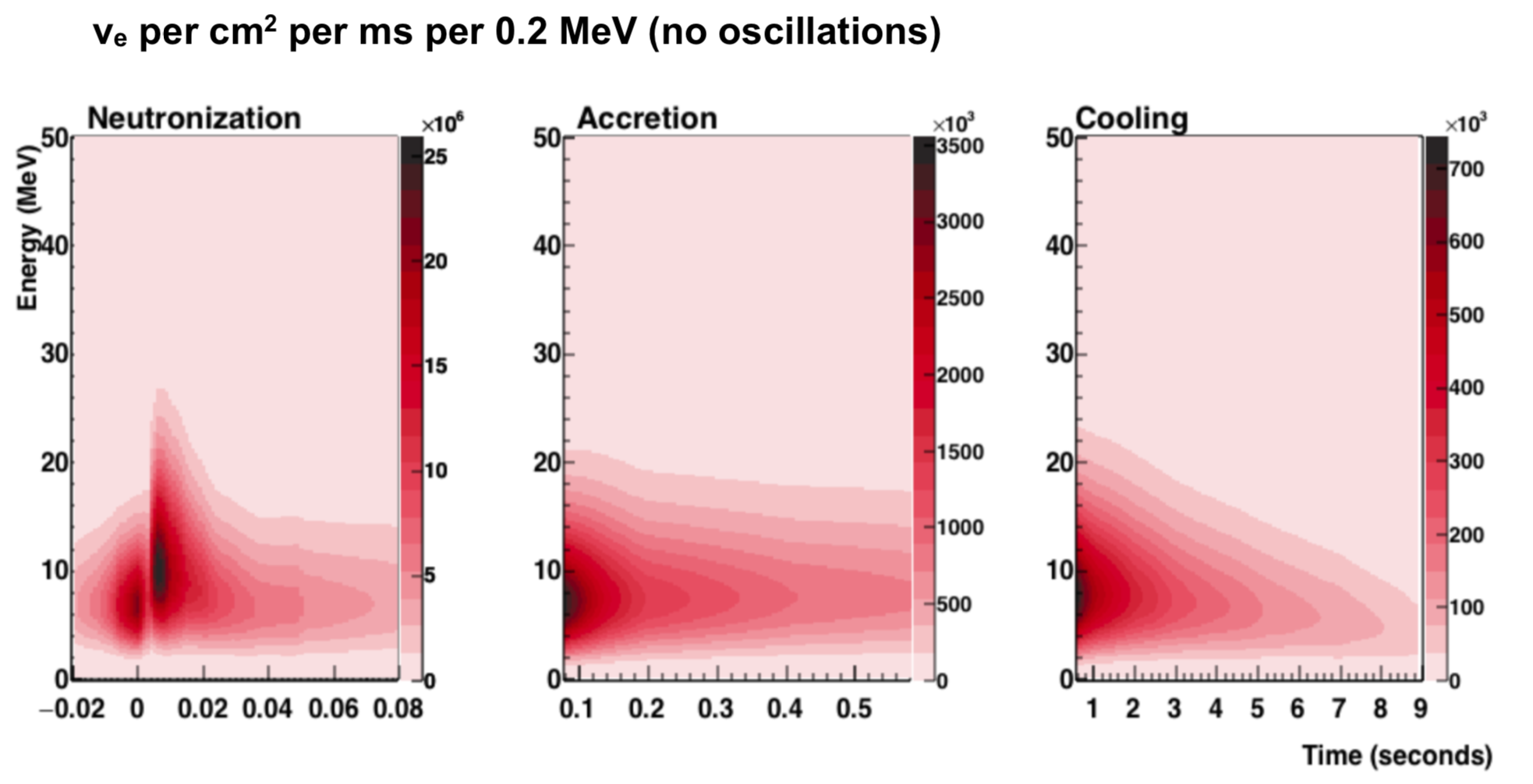}
\caption{Electron neutrino flux \cite{bib3} for the electron-capture supernova model in Ref.~\cite{Huedepohl}.}
\label{fig:flux}
\end{figure}

Neutrinos of all flavors are emitted (Figure \ref{fig:cross}) but thanks to the underground location of the FD, DUNE will be extraordinarily sensitive to the electron neutrino component (Figure \ref{fig:flux}), via the charged-current absorption on $^{40}$Ar (dominant interaction): \\

\vspace{-0.5cm}

\begin{equation}
\nu_{\text{e}} + ^{40}\text{Ar} \rightarrow \text{e}^{-} + ^{40}\text{K}^{*}
\end{equation}

\section{Recent progress on detection and reconstruction of SN neutrinos in DUNE}

Different studies on optimization of the SN detection capabilities of DUNE are being performed by the DUNE SNB/LE Physics Working Group \cite{bib3}. Some of them are commented on this section. In general, SN neutrinos have energies of $\sim$5-50 MeV, much lower than the energies of the neutrinos from the beam ($\sim$GeV). Therefore, special reconstruction algorithms must be developed for that challenging low-energy regime.

\begin{figure}[h]
\centering
\includegraphics[height=0.28\textwidth]{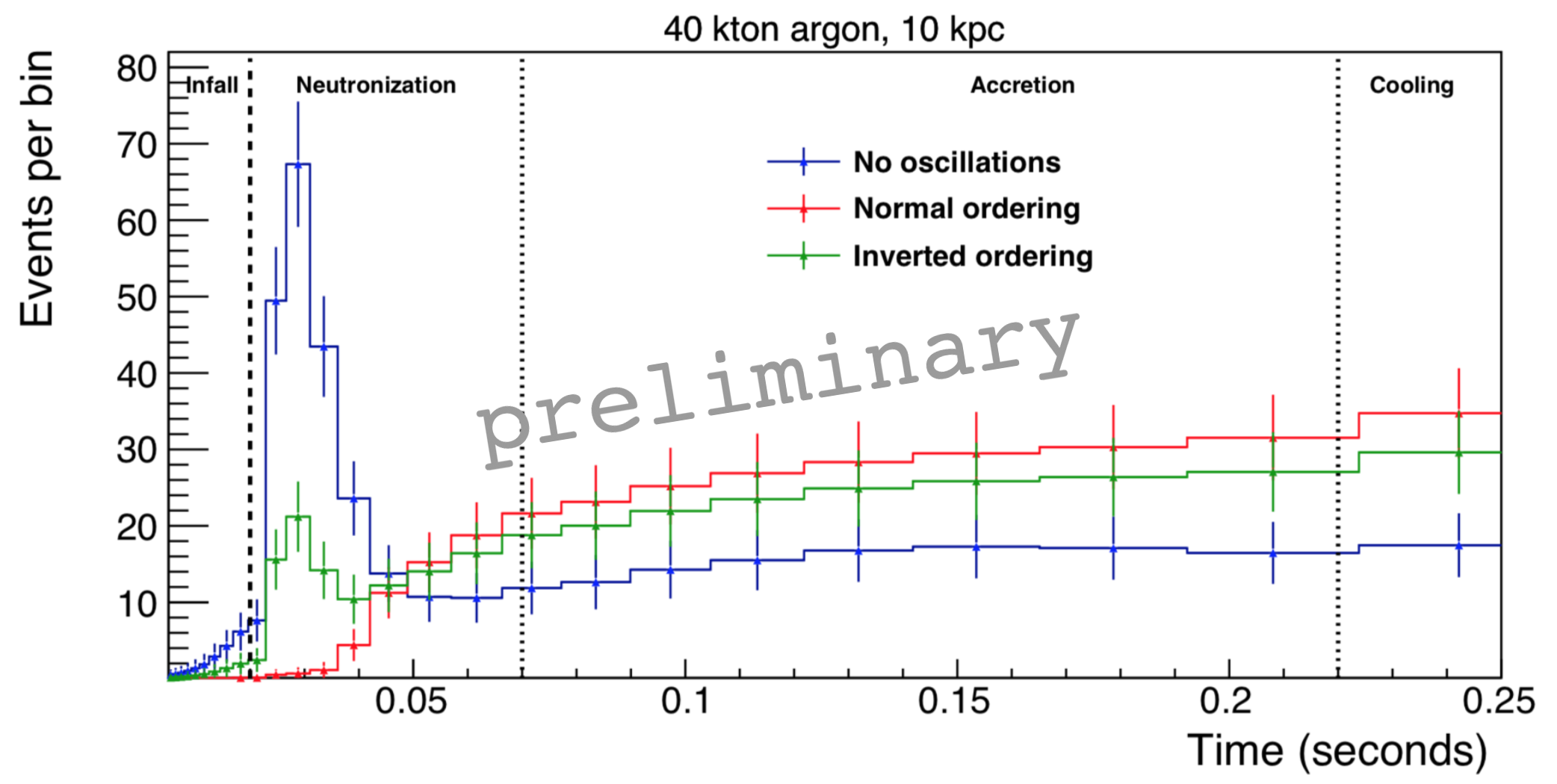}
\includegraphics[height=0.28\textwidth]{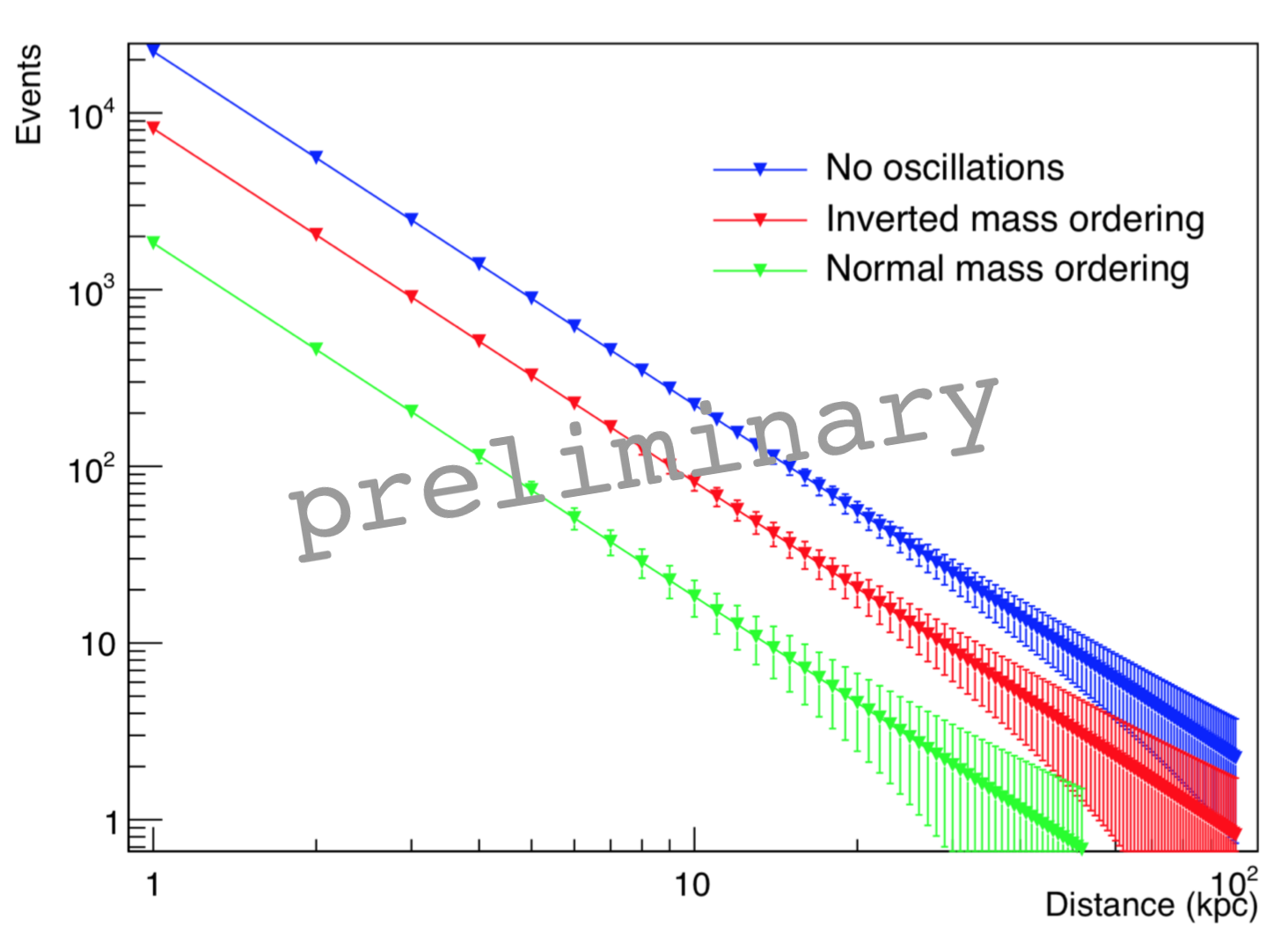}
\includegraphics[height=0.28\textwidth]{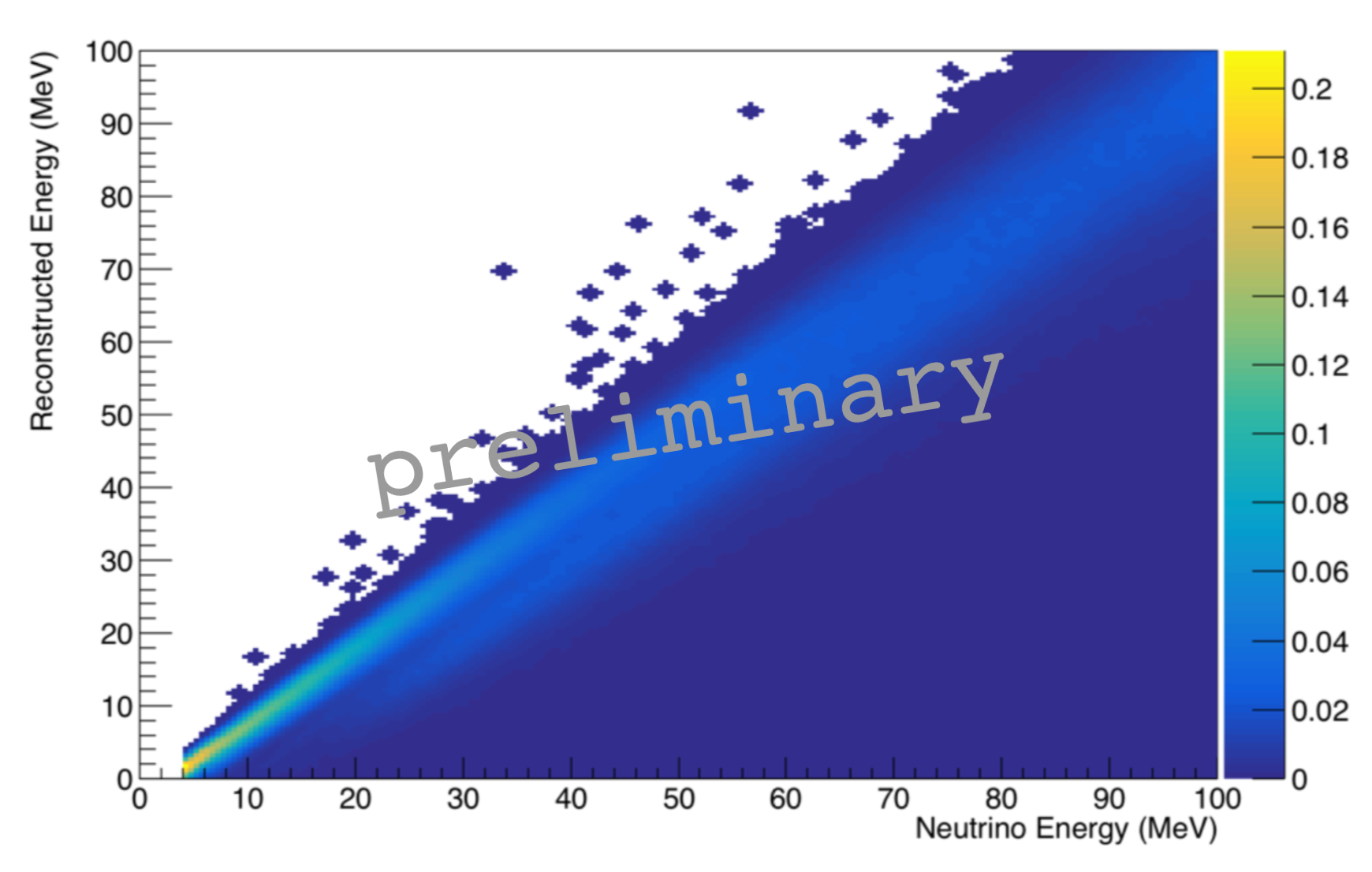}
\includegraphics[height=0.28\textwidth]{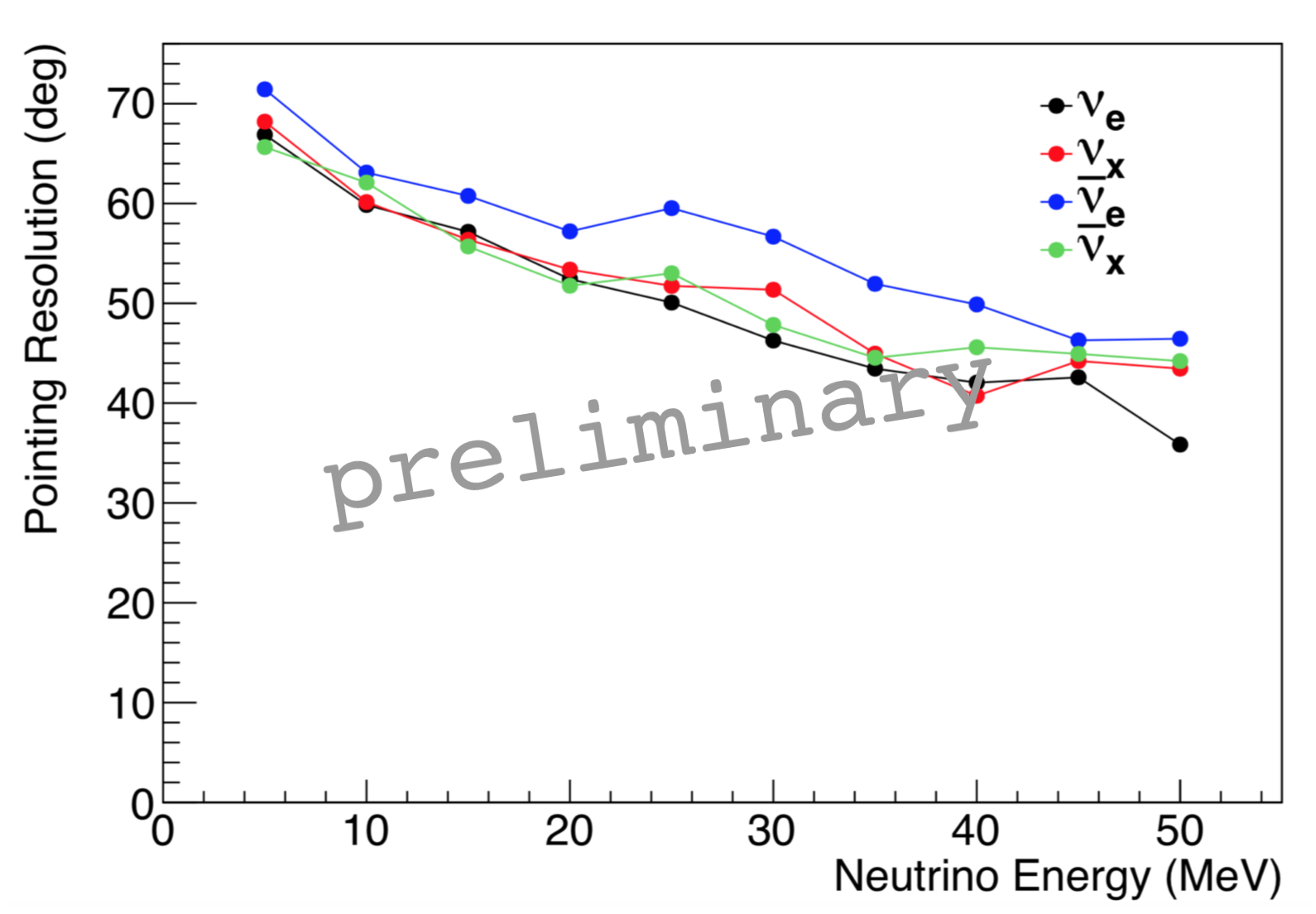}
\caption{\textbf{Top left:} Event distribution during the first 250 ms of the burst for the SN model in Ref.~\cite{Huedepohl}. \textbf{Top right:} Event counts in the first 50 ms as a function of distance under different mass ordering assumptions (statistical errors). \textbf{Bottom left:} Smearing (transfer) matrix including drift correction from the light. \textbf{Bottom right:} Pointing resolution study for elastic scattering events.}
\label{fig:group}
\end{figure}

The mass ordering discrimination in DUNE using the neutronization burst signal has been investigated (Figure \ref{fig:group} top left). The neutronization peak is strongly suppressed for normal ordering and suppressed but still visible for the inverted one, leading to a clear discrimination. 

Calculations of the expected event counts in the neutronization burst as a function of the distance under different mass ordering assumptions are shown in Figure \ref{fig:group} top right, for the specific supernova flux model in Ref.~\cite{Huedepohl}. The rates are well separated out to the edge of the Galaxy for this particular model. Other models are under investigation.

The smearing matrix (reconstructed versus true neutrino energy) has been obtained from SN$\nu_{\text{e}}$CC samples (Figure \ref{fig:group} bottom left). It includes a drift correction using the timing from optical photon detection. This matrix is used, for example, to calculate the expected event rates with SNOwGLoBES \cite{bib5}. 

DUNE will be able to reconstruct the direction of the incoming neutrinos and this will be really useful for a prompt detection of the SN light. In particular, a pointing resolution study of elastic scattering events can be seen in Figure \ref{fig:group} bottom right. In this study, the absolute values of the cosines of the angular differences (neutrino-electron) are used. For an ensemble of elastic scatters from a supernova, the overall pointing resolution will be $\sim$5$^\circ$; studies in the presence of background are underway.

\section{SN trigger using the Dual Phase Photon Detection System}

The photon detection system (PDS) of DUNE will be able to provide a trigger to detect SN neutrinos  \cite{bib6,bib8}. The combination of both the charge and light informations is essential for a redundant and highly efficient trigger scheme in DUNE. The DP PDS will have 720 TPB coated Hamamatsu R5912-MOD20 PMTs uniformly distributed on the floor of the detector. 

The SN$\nu_{\text{e}}$CC event generation (MARLEY \cite{bib7}), propagation (Geant4) and detection have been simulated in LArSoft\footnote{Liquid Argon Software (LArSoft) is a shared base of physics software across LArTPC experiments.}. The events are uniformly generated over the TPC active volume. A sample of radiological background events has been obtained with $^{39}$Ar (1.01 Bq/kg) and neutrons ($10^{-5}$ cm$^{-2}$ s$^{-1}$) as main components. The PMT response has been integrated in LArSoft. The expected light yield at the nominal electric field (500 V/cm) is $\sim$24k photons/MeV. Different configurations of reflector/wavelength shifting panels on the inner walls of the TPC have been considered: 1) no-foils, 2) half-foils (only on the top half of the detector) and 3) full-foils. Thanks to the foils, the light yield reaching the PMTs from the top part of the active volume increases. 

\begin{figure}[h!]
\centering
\includegraphics[height=0.35\textwidth]{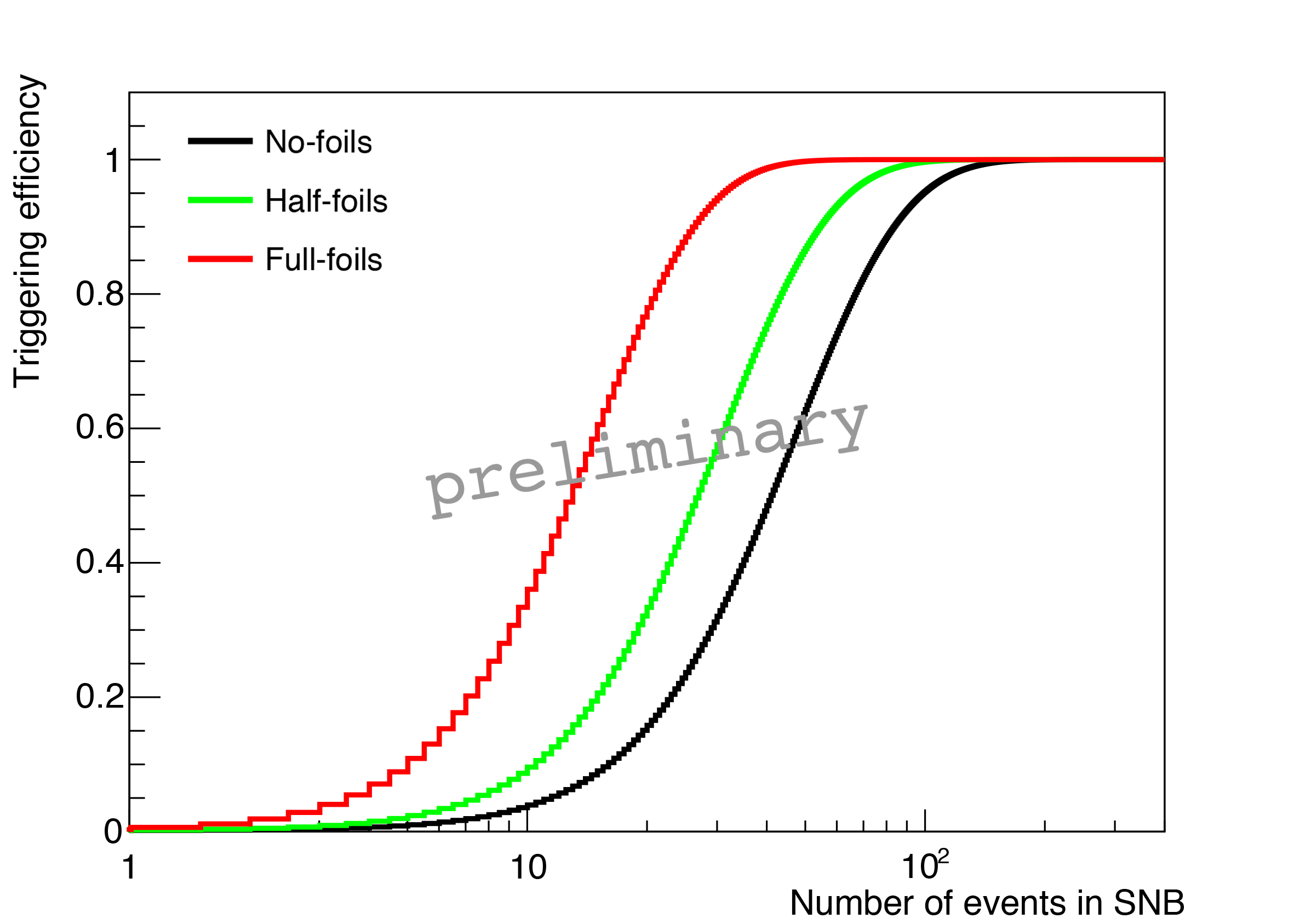}
\includegraphics[height=0.35\textwidth]{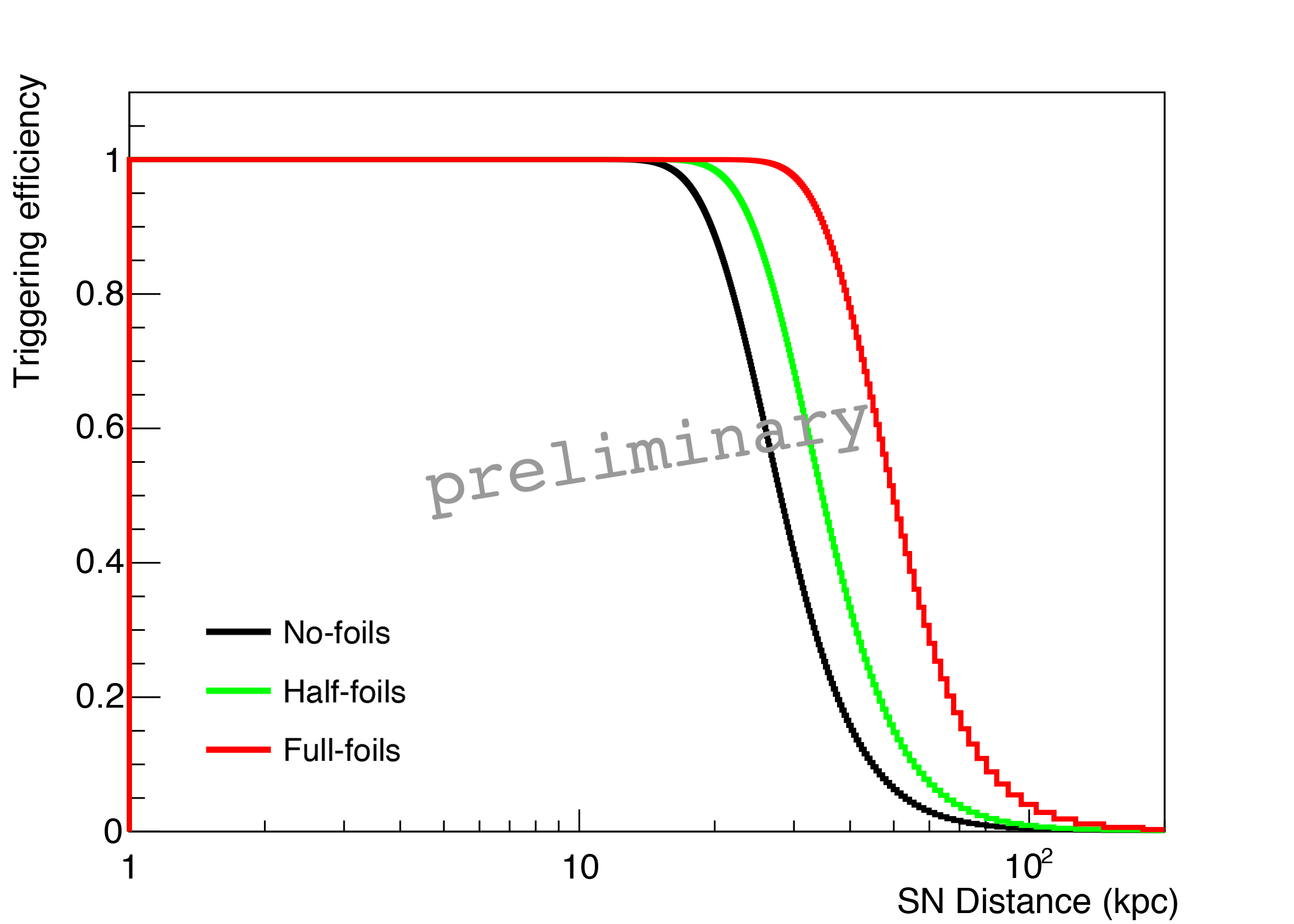}
\caption{\textbf{Left:} SN burst triggering efficiency as a function of the number of events in the SN burst. \textbf{Right:} SN burst triggering efficiency as a function of the SN distance.  }
\label{fig:pmts}
\end{figure}

First, the light signals or hits in the PMTs are identified as waveform pulses. Second, the optical hits are grouped in clusters. The definition of optical cluster is optimized to minimize the detected background rate while increasing the SN signal detection efficiency. Finally, the triggering efficiency is obtained as a function of the number of SN events in the burst (Figure \ref{fig:pmts} left) and the SN distance (Figure \ref{fig:pmts} right) for the best clustering definition. The clustering parameters to be optimized are: 1) minimum number of hits, 2) maximum hit distance, 3) maximum time between hits and 4) maximum cluster duration. The half-foil configuration has been adopted as new baseline for the DP detector because a good triggering efficiency (>90\%) is reached up to 24 kpc and it also allows to fulfil the requirements for proton decay studies.

\section{Conclusions}
One of the key science goals of DUNE is the detection of the next SN neutrino burst in our Galaxy. Different studies are being developed to understand and optimize the SN detection capabilities of DUNE.  Regarding the SN trigger, the simulation studies show that the DP PDS will yield a highly efficient trigger for a SN burst occurring anywhere in the Milky Way.



\end{document}